\documentclass[prl,aps,twocolumn,showpacs]{revtex4}
\usepackage{epsfig,amssymb,amsfonts}
\newcommand\beq{\begin{equation}}
\newcommand\eeq{\end{equation}}
\newcommand\bea{\begin{eqnarray}}
\newcommand\eea{\end{eqnarray}}

\newcommand\stm{\textsf{STM~}}
\newcommand\stmd{\textsf{STM}}

\newcommand\tll{\textsf{LL~}}
\newcommand\tlld{\textsf{LL}}

\newcommand\sv{\textsf{SV~}}
\newcommand\svd{\textsf{SV}}
\newcommand\hll{\textsf{HLL~}}
\newcommand\hlld{\textsf{HLL}}

\def\lsim{\mathrel{\rlap{\lower4pt\hbox{\hskip1pt$\sim$}}
\raise1pt\hbox{$<$}}} 
\def\gsim{\mathrel{\rlap{\lower4pt\hbox{\hskip1pt$\sim$}}
\raise1pt\hbox{$>$}}} 
\def\dfrac#1#2{{\displaystyle\frac{#1}{#2}}}
\newcommand\odd{1{\textendash}D}

\newcommand\ie{{\textit{i.e.}}}

\begin{document}
\textheight=23.8cm

\title{\Large Spin polarised scanning tunneling probe for \\
helical Luttinger liquids
}
\author{\bf  Sourin Das$^{1,2,3}$ and Sumathi Rao$^{4,5}$ }
\affiliation{\it$^1$  Department of Physics and Astrophysics, University of Delhi,
Delhi 110 007, India\\
\it $^2$ Institut f\"{u}r Festk\"{o}rper-Forschung -
Theorie 3 Forschungszentrum J\"{u}lich, 52425  J\"{u}lich, Germany \\
$^3$ Institut f\"{u}r Theoretische Physik A, RWTH
 Aachen, 52056 Aachen, Germany\\
 $^4$ LPTHE, Universite Pierre et Marie Curie, Paris 6,
4, Place Jussieu, 75252, Cedex 05, France \\
 $^5$ Harish-Chandra Research Institute, Chhatnag Road, Jhusi,
Allahabad 211 019, India}
\date{\today}
\pacs{71.10.Pm, 71.27.+a, 73.40.Gk}

\begin{abstract}

We propose a three terminal spin polarized \stm setup for probing
the helical nature of the
Luttinger liquid  edge state that appears in the quantum spin Hall system.
 We show that the
three-terminal tunneling conductance strongly depends on the angle ($\theta$)
 between the
magnetization direction of the tip and the local orientation of the electron
 spin on the edge
while the two terminal conductance is independent of this angle.
 We demonstrate that chiral
injection of an electron  into the helical Luttinger liquid (which
occurs when $\theta$ is
zero or $\pi$) is associated with fractionalization of the spin of
the injected electron in
addition to the fractionalization of its charge. We also point out a
spin current amplification
effect induced by the spin fractionalization.

\end{abstract}

\maketitle

\noindent {\it Introduction :-}
A new class of insulators have recently emerged called quantum
spin Hall insulators which have gapless edge states due to the
topological properties of the band structure \cite{kane}. For a
two-dimensional insulator, a pair of one-dimensional counter
propagating modes  appear on the edges \cite{kane,buttiker} which
are  transformed into helical Luttinger liquids (\hlld) due to
inter-mode Coulomb interactions \cite{xu}.  Various  aspects of this
state\cite{wu,chamon,oreg,nagaosa,anders,joel} have been studied.
The central point about
the \hll is the fact that  the  spin orientation of the edge
electrons, which is dictated by the bulk physics, is correlated with
the direction of motion of the electron - $i.e.$, opposite spin
modes counter propagate. The existence of such edge channels have
already been detected experimentally in a multi-terminal Hall bar
setup \cite{roth}. But although this  experiment does confirm the
existence of counter propagating one-dimensional (\odd) modes at the
edge, it is not a direct observation of the spin degree of freedom.
A central motivation of this letter is to suggest  a setup wherein
the structure of the spin degree of freedom on the edge can be
directly probed.

Motivated by the spin valve (\svd) effect, the first idea to probe
the spin degree of freedom,
would be to replace one of the  ferromagnetic leads in a magnetic tunnel
junction by the \hll
and measure the magneto-resistance, as a function of the
relative spin orientation of the \hll
and the magnetization direction of the ferromagnetic lead.
However,  the angle dependent tunnel
resistance for the \sv depends directly on the degree of
polarization of the two leads. For \hlld,
although the edge modes have a specific spin orientation locally,
 they have no net polarization,
and hence the tunnel resistance would be independent of the
spin polarization of the ferromagnetic
lead.

In this letter, we show that switching  to a three terminal geometry
involving a magnetized scanning tunneling microscope (\stmd) tip
facilitates the detection of the spin orientation of the edge
electron by inducing a finite three terminal magneto-resistance. For
a normal \tlld, it is not possible to inject an  electron with a
well-defined momentum (left or right movers) at a localized point in
the wire, and  hence extended wires were used as injectors in
Ref.~\onlinecite{Steinberg} to achieve chiral injection. But for
\hlld,  since the direction of motion is correlated with the spin
projection, chiral injection (\ie, injecting only left movers or
right movers)  is possible even at a localized point in the wire. One
just needs to tune the direction of polarization of the \stm
parallel (anti-parallel) to the polarization direction of the edge.
Once this is achieved, injection of an upspin (downspin) electron is
equivalent to injecting right (left) movers. Hence the \hll  has a
natural advantage over a normal \tll for chiral injection. As was
experimentally demonstrated in Ref.~\onlinecite{Steinberg}, chiral
injection of electrons can lead to an  asymmetry in the currents
measured on both sides of the injection region, which is further
modified by the \tll interaction. In this letter, we show that in a
similar setup, the left-right current asymmetry in the wire when
voltage biased with respect to the \stm  has also  a strong $\theta$
dependence due to the interaction induced scattering of electrons
between the right (spin up) and left (spin down) moving edges. For
purely chiral injection ($\theta=0,\pi$),  we find that the fraction
of the total tunneling current measured at the left and right of the
injection region is  asymmetric and is given by the splitting
factor (left) $A_{c1}=(1\mp K)/2$ and (right) $A_{c2}=(1\pm K)/2$
(where K is the \tll parameter and the top and bottom signs are for
$\theta=0$ and $\pi$ respectively) in agreement with the results in
Ref.~\onlinecite{hur},   and is a manifestation of charge
fractionalization of the injected electron.
Observing the
asymmetry with a spin polarized \stm as a local injector would be an
indisputable sign of the helical nature of the edge states, since
for the usual \tlld, no such current asymmetry would be expected for
local injection.

The most subtle outcome of our analysis is the fractionalization
of the spin of the injected electron. In
contrast to  charge fractionalization,  the spin gets fractionalized
such that one of the fractionalized
components turns out to be larger than the injected spin.
The asymmetric fractions of the total injected electron spin current
are  given by $A_{s1}=(1 \mp K)/4K$
and $A_{s2}=(1 \pm K)/4K$ (upper
and lower signs for $\theta=0$ and $\theta=\pi$ respectively) and are
a manifestation of the fractionalization
of the injected electron spin in the \hlld. Note that for $K<1$
(repulsive electrons) $A_{s2}>1/2$, thus
resulting in an effective magnification of the injected spin current
 at the right lead.

\begin{figure}
\begin{center}
\includegraphics[width=5cm,height=4cm]{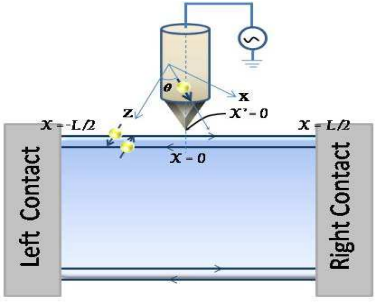}
\caption{A schematic of the geometry of the proposed setup.
The direction of orientation of the electron spin in the \hll
is along the  ${\widehat Z}$ axis. The angle between direction of
orientation of the spin of electrons in the edge and the majority
spin in the \stm tip is $\theta$  and they are assumed to lie in the
${\widehat X}$-${\widehat Z}$ plane. The ${\widehat Y}$-axis points
out of the plane of the paper. Here $x$ and $x^\prime$ represent the
intrinsic one dimensional coordinates
 of the \stm tip and the wire.
 }
\label{figone}
\end{center}
\end{figure}
 \noindent
{\it Geometry :-}
We propose a three terminal junction as shown in Fig.~\ref{figone}.
Three terminal setups have also been used to study tunneling
into a quantum wire in the Fabry-Perot regime\cite{dolcini}.
The spin of the electrons in the edge states are polarized in some
direction depending on details of the spin-orbit interaction in the
bulk. We use a coordinate system which has its ${\widehat Z}$-axis
along the direction of orientation of the spin of the edge electrons
and the plane containing the polarization direction of the edge
electron and the tip electron is assumed to be the ${\widehat
X}$-${\widehat Z}$ plane (see Fig.~\ref{figone}) . Note that here we
have assumed that the edge is  smooth and is along a
straight line, so that there is a well defined quantization
direction for the electron spin living on the edge.

The Hamiltonian for the \hll is given by \cite{chamon}
\beq H_0 = v \int_{-L/2}^{L/2} dx~ \left[~K (\partial_x \Phi)^2 +
K^{-1} (\partial_x \Theta)^2~\right]~, \label{one} \eeq
where $\Phi=(\phi_{R\uparrow} + \phi_{L\downarrow})/2$,
$\Theta=(\phi_{R\uparrow} - \phi_
{L\downarrow})/2$ and the  $\phi_{R\uparrow/L \downarrow}$
are related to the up and down
electron operators  in the edge by the standard bosonization identity
$\psi_{R\uparrow}(x)
\sim \frac{1}{\sqrt{2\pi\zeta}}e^{ik_Fx} e^{i\phi_{R\uparrow}(x)},
\psi_{L\downarrow}(x) \sim
\frac{1}{\sqrt{2\pi\zeta}}e^{-ik_Fx} e^{i\phi_{L\downarrow}(x)}$.
$\zeta$ and $K$ are
the short
distance cut-off and the Luttinger parameter respectively.
Unlike  the
standard \tlld, here the spin orientation is correlated with
the direction of motion. We
drop  Klein factors as they are irrelevant for our computations.

The Hamiltonian for the \stm is assumed to be that of a free
electron in \odd. The tunneling Hamiltonian between the tip and the
helical edge at a position $x=0$, $x'=0$ is given by
\beq H_t = t   ~[~\psi_{i\alpha}^\dagger(x=0) ~\chi_{\alpha}^{}(x'=0)
+ {\rm{h.c.}}~]~, \label{tunnel} \eeq
where $i=R,L$ denotes right and left movers and $\alpha$
denotes the spin index, $\psi_ {i\alpha}$ and $\chi_{\alpha}$ denote
the electron destruction operator in the \hll and the \stm
respectively.
Voltage bias in the tunneling operator can be introduced simply by
replacing $ \chi_ {\alpha}^{}(x) \rightarrow \chi_{\alpha}^{}(x)
e^{-iVt/\hbar}$. We will, henceforth, drop the index $i,j$ denoting
the direction of motion.

Since the tunneling conserves spin, using a fully polarized \stm
with polarization direction tuned along the positive or negative
direction of  ${\widehat Z}$-axis will naturally allow for chiral
injection \ie, injecting only right ($\uparrow$) or left
($\downarrow$) movers. In the absence of interactions in the \hlld,
the chirally injected electron will cause both charge current and
spin current to flow only to the right or to the left lead,  hence
leading to a left-right asymmetry. In the presence of interactions
in the \hlld, due to Coulomb scattering between the right and left
movers, the chirally injected charge and spin degrees of freedom of
the electron get fractionalized and  move in both directions;
however, in general, the left-right asymmetry  still survives.

Now, let us consider the fully polarized \stm tip with the
polarization direction making an arbitrary angle $\theta$ with
respect to the spin of the \hll electron. In the quantization basis
of the \hll spins, the  tip spinor can be written as
$ \chi_{\rm{rot}}^{} = e^{-{i\theta{\sigma}\cdot{\hat Y}}/{2}}~
\chi_T^{}~, $
where $\chi_T^{}$ is the tip spinor in a basis where the spin
quantization axis is along the \stm polarization direction $i.e.$,
$\chi_T^{} = (\chi_{\uparrow}^{}, 0)$. So $\chi_{{\rm{rot}}
\uparrow}^{} = \cos(\theta/2)~ \chi_{\uparrow}^{} $ and
$\chi_{\rm{rot} \downarrow}^{} = \sin(\theta/2)~
\chi_{\uparrow}^{}$. In other words, the electron in the tip has
both $\uparrow$ and $\downarrow$ spins,
but the effective tunnel amplitudes
are asymmetric (except when $\theta=\pi/2$) and hence, the current
asymmetry survives. As a function of the rotation angle $\theta$,
the chiral injection goes from being a pure right-mover at
$\theta=0$ to a pure left mover at $\theta=\pi$.

 \noindent {\it Charge current :-}
The tunneling Hamiltonian  can now be rewritten in terms of
$\chi_{\uparrow}^{}$ as
\beq H_t = \left[~t_{\uparrow}^{} ~ \psi_{\uparrow}^\dagger ~
\chi_{\uparrow}^{} + t_{\downarrow}^{} ~ \psi_{\downarrow}^\dagger~
\chi_{\uparrow}^{} + {\rm{h.c.}}~\right]~, \label{tunnel2} \eeq
where $t_\uparrow^{} = t \cos(\theta/2)$ and $t_\downarrow^{} = t
\sin(\theta/2)$ can be tuned by tuning $\theta$. The Boguliobov
fields $\widetilde \phi_{L/R}^{}$ which move unhindered to right and
left direction (henceforth we call them the right chiral and left
chiral fields)  are given by
\bea \phi_{\uparrow/\downarrow}^{} &=& \dfrac{1}{2 {\sqrt
K}}~\left[~(1 \pm K)~ {\widetilde \phi_R} + (1 \mp K) ~{\widetilde
\phi_L}~\right]~. \eea
Note that the total electron density on the \hll  wire can be
expressed in terms of the chiral fields as $\rho (x) =
({\sqrt{K}}/{2 \pi}) \partial_x \widetilde \phi_R^{} -
({\sqrt{K}}/{2 \pi})
\partial_x \widetilde \phi_L^{}$ thus defining the chiral right
(left) densities and the corresponding number operators as
\beq \widetilde N_{R/L} = \int_{-L/2}^{L/2} dx ~\widetilde
\rho(x)_{R/L}^{} = \pm  \dfrac{\sqrt{K}}{2 \pi} \int_{-L/2}^{L/2}
dx~
\partial_x (\widetilde \phi_{R/L})~. \label{cdensity} \eeq
Next we define the operator corresponding to the chiral decomposition
of the total charge current as
$ I_{t\alpha} = {d{\widetilde N}_\alpha}/{dt} =
-i[{\widetilde N}_\alpha,H_t], $
where we have set $\hbar=1$ and electron charge $e=1$ and
$\alpha={R/L}$. Using the standard commutation relations of chiral
fields, $[{\widetilde \phi}_{\uparrow/\downarrow}(x), {\widetilde
\phi}_{\uparrow / \downarrow}(x')] = \pm  i\,\pi \,{\rm sgn} (x-x')$
the chiral currents  can be found to be
\bea
I_{t{R/L}}^{}(\theta)& = &   \dfrac{1}{2} [~(1 \pm K)~\cos ({\theta}/{2})
~I_{t}^{} (\theta=0)  \nonumber \\
                   &+& (1 \mp K)~ \sin ({\theta}/{2}) ~ I_{t}^{}
(\theta=\pi) ~]~.
\label{ccurrent}
\eea
$I_{t} (\theta)= I_{tL} (\theta)+ I_{tR}(\theta)$ is the total
tunneling charge current operator for an arbitrary value of $\theta$
and $I_{t}(\theta=0/\pi) = i t (\chi_{\uparrow}^\dagger
\psi_{\uparrow/\downarrow}^{} -\psi_{\uparrow/\downarrow}^\dagger
\chi^{}_{\uparrow}$). The expectation values of the currents
operator in linear response is given by
\beq \langle ~I_{t}^{}(\theta)~ \rangle =-\dfrac{i}{\hbar}
\int_{-\infty}^{0} d \tau ~\langle~[~I_{t}(\theta,\tau=0),
H_t(\tau)~]~\rangle~. \eeq
Since the \hll Hamiltonian is left-right symmetric in the absence of
the tip and  the tip is fully polarised, the value is equal for  $\theta=0$
and $\theta = \pi$
and given by
$ \langle~I_t(\theta=0) ~\rangle =
\langle~I_t(\theta=\pi)~\rangle = I_0~. $
Using the well-known correlation function of \tll liquid at finite
temperature $T$, we find
\beq I_0 = {\dfrac{e^2}{h}} \vert t^2 \vert  \dfrac{(T/\Lambda)^\nu}
{(\hbar v_F)^2 ~\Gamma (\nu+1) } \times V~, \label{currenti0} \eeq
where $\Lambda$ is an ultra-violet cutoff and $\nu$ is the Luttinger
tunneling exponent given by $\nu=-1 +(K + K^{-1})/2$.  Here we have
have assumed that $T \gg T_L, T_V$, where $T_L$ is  the temperature
equivalent of the length of the wire defined by $v/L=k_B T$ and $T_V
= eV/k_B$, is the temperature equivalent of bias voltage.

Using these values, we now obtain the current heading to the right
and left ends of the wire as
\bea \langle~ I_{tR,L} (\theta) ~\rangle & =&  \dfrac{(1 \pm
K\cos\theta)}{2} I_0~. \label{current} \eea
Note that even though the left and right chiral currents which will
be measured at the right and left contact depend on $\theta$,  the
total tunneling current $I_{t} (\theta)= I_{tL} (\theta)+
I_{tR}(\theta)$ is independent of $\theta$.  Thus we show
that unlike the  two terminal tunnel current,
 the three terminal current is clearly not independent
of $\theta$. This is one of the key results of this letter.

\noindent {\it Spin currents :-}
The isolated \hlld,  even in equilibrium,  has a persistent spin
current because of the correlation of the direction of spin with the
direction of motion but no charge current. However, here we would
like compute the excess spin current that is caused by the inflow
of electrons from the \stm tip into the edge mode. Now the tunneling
induced magnetization of the edge state can be defined as ${\bf
S}={\int_{-L/2}^{L/2} dx~ {\bf s}(x)} = \int_{-L/2}^{L/2} dx~
(\psi_\alpha^\dagger {\vec \sigma}_{\alpha\beta} \psi_{\beta}/2)$
where ${\bf s}(x)$ is the local spin density. Hence the spin current
can be defined as $d{\bf S}/dt=-i [{\bf S},H_t] $. Now using
bosonization,  it is straight-forward to evaluate the ${\widehat
X}$, ${\widehat Y}$ and ${\widehat Z}$ components of the spin
current operator as given below
\bea {\dot{S_X}(\theta)}  &=& \dfrac{1}{2} \left[
\cos(\theta/2)I_{t}(\theta=\pi) +
\sin(\theta/2) I_{t}(\theta=0) \right]~, \nonumber\\
{\dot{S_Y}(\theta)}  &=& \dfrac{1}{2}
\left[ \cos(\theta/2)H_{t}(\theta=\pi) - \sin(\theta/2) H_{t}(\theta=0)
\right]~, \nonumber\\
{\dot{S_Z}(\theta)}  &=& \dfrac{1}{2K} \left[I_{tR}(\theta) -
I_{tL}(\theta)\right]~. \label{spin1} \eea
Note that ${\dot{S_X}}$ and ${\dot{S_Z}}$ are expressible in terms
of the current operator while the ${\dot{S_Y}}$ is expressible only
in terms of the tunnel Hamiltonian given in Eq.~\ref{tunnel2}. The
difference is related to the fact that only the ${\widehat X}$ and
${\widehat Z}$ components  of the spin  are relevant as the injected
electron spin from
 the \stm has no component along the ${\widehat Y}$ direction.  Hence
$\dot{S_Y}$ is expected to be zero and
indeed the expectation value of ${\dot {S_Y}}$ is easily seen
to be zero,  since  $H_t$ is left-right symmetric.
Now using Eqs.~\ref{ccurrent},~\ref{current} and \ref{spin1},
 we get the following expressions (within linear response)
 for the spin currents towards the left and right contacts -
\beq {{\bf \dot{S}}_{R/L}(\theta)} = \left[~ \dfrac{K \mp
\cos\theta}{2 K \sin\theta}~{\widehat Z} ~\pm~ \frac{1}{2
K}~{\widehat{X}} ~\right] ~I_{t{R/L}}(\theta)~. \label{spin2} \eeq
Hence,  for
arbitrary values of  $\theta$, the spin current collected at the right and the
left contacts are asymmetric.  Now using Eqs.~\ref{current} and \ref{spin2},
it is easy to check
 that total injected spin current
\beq \langle~ \dfrac{d{\bf S}}{dt}~ \rangle
=(~{\widehat{Z}}\cos\theta +
{\widehat{X}}\sin\theta~)~\dfrac{I_0}{2}~, \label{spin3} \eeq 
is pointing exactly along the magnetization
direction of \stm as expected.

\noindent {\it Charge and spin fractionalization :-}
 Recently, the issue of
charge fractionalization has been addressed  both theoretically and
experimentally
 in Refs.~{\onlinecite{Steinberg,hur}} and {\onlinecite{dolcini}}.
 The fractionalization of
a chirally injected electron charge
into the \hll  at a point ($x=0$) can be understood by considering the
following commutator
\beq \left[\widetilde \rho_{R/L}^{} (x), \psi_{R}^\dagger(0)\right]
= \dfrac{1 \pm K}{2}~ \delta (x)~\psi_{R}^\dagger(0)~. \eeq
This implies that the creation of a single right moving electron at $x=0$
creates simultaneously
an excitation of charge $(1 \pm K)/2$  in the right and left going chiral
densities, thus leading to fractionalization of electron. (A similar
equation (with an overall sign change) works for the left-movers).
Note that the splitting of the total tunneling current
 into its chiral components (see Eq.~\ref{ccurrent}) is
exactly consistent with the splitting
of the electron charge.
Hence measuring the
chiral currents can provide information about the charge fractionalization, as
 demonstrated in Ref.~\onlinecite{Steinberg}.

Similarly, to study spin fractionalization, we bosonize the
${\widehat Z}$-component of spin density given by $s_Z (x) = (1/2)
(\psi_{\uparrow}^\dagger(x) \psi^{}_{\uparrow}(x)
-\psi_{\downarrow}^\dagger(x) \psi^{}_{\downarrow}(x))$ to obtain
$s_Z(x) = (1/2 K) (\widetilde{\rho}_R
(x)-{\widetilde{\rho}_L (x)})$.
This defines  $s_{Z,R/L}=\pm(1/2K) \widetilde{\rho}_{R/L} (x)$. Now
let us consider the following commutator
\beq \left[~s_{Z,R/L}(x), \psi_{R}^\dagger(0)~\right]=\dfrac{1}{2}
~\left(\dfrac{1 \pm K}{2 K}\right)~ \delta (x)~\psi_{R}^\dagger(0)~.
\eeq 
This implies that the creation of a single right moving electron at
$x=0$ creates simultaneously spin excitations of spin $(1 \pm K)/2
K$ (in units of electronic spin quanta) in the right and left going
chiral spin densities, thus  leading to $K$ dependent
fractionalization of the spin of the injected electron. Now let us
consider the ${\widehat Z}$-component of the spin current operator
given in Eq.~\ref{spin1}. This operator can be chirally  decomposed
as follows - 
\beq
\langle {\dot{S_Z}(\theta)}\rangle_{R/L}  = \pm\dfrac{\langle I_{tR/L}
(\theta)\rangle}{2 K}
=\pm\left( \dfrac{1 \pm  K \cos\theta}{2 K}\right) \dfrac{I_0 }{2}.
\eeq
For chiral injection (\ie, $\theta=0,\pi$) we note the splitting of the total
tunneling spin current ($I_0/2$) into its  chiral components is given
precisely by $(1 \pm K)/{2 K}$,  which is
exactly consistent with the splitting  of the electron spin evaluated
from the commutator.
Intriguingly, one of the splitting fractions, $(1 + K )/2 K$ is larger than
unity for $K<1$ (\ie, for repulsive electrons). Hence in the three
terminal geometry one obtains an  interaction ($K \neq 1$) induced
amplification of the injected spin current.

\noindent {\it Discussion :-} Regarding the application of our work to realistic systems, we first point out
that our work is directly applicable  to edge states in graphene with a small spin-orbit coupling\cite{kanemele}
and to  other genuine quantum spin Hall insulators like the model considered in Ref.\,\onlinecite{zhabern} and
Bi\,\cite{muruk}. However, for  $HgTe/CdTe$ quantum wells, where the spin projections actually refer to
pseudospin  related to the two block diagonal parts of the effective Hamiltonian written in the $|E_1,m_J=+1/2
\rangle$, $|H_1,m_J=+3/2 \rangle$, $|E_1,m_J=-1/2 \rangle$ and $|H_1,m_J=3/2 \rangle$ basis\cite{Koenig}, we
need to modify our computations. Fig.~1 is still applicable with the ${\widehat Z}$-axis now referring to the
crystal growth axis of the quantum well. But the  $\psi_{\uparrow/\downarrow}$ states  that we have considered
in Eq.\ref{tunnel2} are no longer right or left movers even before interactions have been introduced. We need to
introduce the right and left moving fields as $\eta_{\uparrow/\downarrow}$ given by $ \psi_\uparrow = a
\eta_\uparrow + b \eta_\downarrow$ and $ \psi_\downarrow = a' \eta_\uparrow + b' \eta_\downarrow $ where
$a,b,a',b'$ are material dependent parameters which denote how the  pseudospin states are related to the real
spin of the electron. Hence, the tunneling Hamiltonian in Eq.\ref{tunnel2} can be rewritten as \beq H_t =
\left[\{(t_{\uparrow}^{} a + t_\downarrow a')~  \eta_\uparrow ^\dagger ~
 + (t_{\uparrow}^{} b+ t_\downarrow b')
\eta_{\downarrow}^\dagger \} \chi_{\uparrow}^{} + {\rm{h.c.}}~\right]~. \label{tunnel3} \eeq We get pure
right-moving or left-moving currents at $\tan\theta/2 = -a'/a$ or at $\tan\theta/2 = -b'/b$. Note that the angle
at which the left-moving current disappears is not exactly opposite to the angle at which the right-moving
current vanishes, since the real spin of the left-movers and right-movers need not be equal and opposite. With
interactions, it is the $\eta_{\uparrow/\downarrow}$ fields which are bosonised and the rest of the formulation
goes through provided that the non-interacting reference angles ($i.e.$,  the coefficients $a,b,a', b'$) are
known. But determining both  $a,b,a', b'$ and $K$  is a non-trivial problem. However, if the experiment could be
carried out at different temperatures at fixed $\theta$, then since the current $I_0$ (defined in
Eq.\ref{currenti0}) depends only on $K$ and not on $\theta$, it may be possible in principle (albeit difficult
in practise) to extract the value of $K$ from the power law dependence of current. Moreover, the edge states
could be known (from other experiments) to be in the weakly interacting regime ($K=1$). In these cases, this
setup can be used  to extract the values of the coefficients $a, b, a', b'$.

 \noindent {\it Conclusion
:-} To summarise, in this letter, we have proposed a three-terminal
polarised \stm set-up as a probe for \hlld. We suggest that the spin
polarized tip can facilitate local chiral injection.
This leads to current asymmetries, with specific
$\theta$  dependence,
whose  measurement can lead to undisputed
confirmation of the helical nature of the edge state.
Chiral injection of the electron into the \hll is
also shown to be directly related to the physics of fractionalization
of the injected electron spin in addition to the fractionalization of
its charge. We
also point out that spin fractionalization leads to
a spin current amplification effect
  in the three terminal geometry.

\acknowledgements{SD would like to  thank  C. Br\"{u}ne, Y. Gefen,
M. Zahid Hasan, M. K\"{o}nig, A. Mirlin, Y. Oreg, G. Refael, K.
Sengupta, S. Simon, M. R. Wegewijs and A. Yacoby for stimulating
discussions. Both of us would like to thank the referee for useful suggestions.}


\vspace{-0.20in} \noindent
\end{document}